\documentclass[aps,prl,reprint]{revtex4-1}
\pdfoutput=1
\usepackage{graphicx}  % needed for figures
\usepackage{dcolumn}   % needed for some tables
\usepackage{bm}        % for math
\usepackage{amssymb}   % for math
\usepackage{color,epsfig}

\def\red#1{\textcolor{red}{#1}}
\def\blue#1{\textcolor{blue}{#1}}

\begin{document}
\newcommand{\HH}{H$_2$}
\newcommand{\pHH}{\emph{para}-H$_2$}
\newcommand{\oHH}{\emph{ortho}-H$_2$}
\newcommand{\Eref}[1]{Eq.~(\ref{#1})}
\newcommand{\Fref}[1]{Fig.~\ref{#1}}
\newcommand{\etal}{\emph{et al.}}

\def\red#1{\textcolor{red}{#1}}
\def\blue#1{\textcolor{blue}{#1}}

% The following information is for internal review, please remove them for submission
\widetext
%\leftline{Version xx as of \today}
%\leftline{Primary authors: ---}
%\leftline{To be submitted to PRL}
%\leftline{Comment to {\tt d0-run2eb-nnn@fnal.gov} by xxx, yyy}
%\centerline{\em D\O\ INTERNAL DOCUMENT -- NOT FOR PUBLIC DISTRIBUTION}

% the following line is for submission, including submission to the arXiv!!
%\hspace{5.2in} \mbox{Fermilab-Pub-04/xxx-E}

%\title{Resonance HD + H$_2$}
\title{Stereodynamical control of a quantum scattering resonance in cold molecular collisions}
\author{Pablo G. Jambrina} \affiliation{Departamento de Qu\'{\i}mica F\'{\i}sica. Universidad de Salamanca, Salamanca 37008, Spain.}\email{pjambrina@usal.es}
\author{James F. E. Croft} \affiliation{The Dood-Walls Centre for Photonic and Quantum Technologies, Dunedin, New Zealand} \affiliation{Department of Physics, University of Otago, Dunedin, New Zealand}\email{j.croft@otago.ac.nz}
\author{Hua Guo}\affiliation{Department of Chemistry and Chemical Biology, University of New Mexico, Albuquerque, New Mexico 87131, United States}\email{hguo@unm.edu}
\author{ Mark Brouard} \affiliation{The Department of Chemistry, University of Oxford, The Chemistry Research Laboratory,
 Oxford OX1 3TA, UK.}\email{mark.brouard@chem.ox.ac.uk}
\author{Naduvalath Balakrishnan}\affiliation{Department of Chemistry and Biochemistry, University of Nevada,
Las Vegas, Nevada 89154, USA}\email{naduvala@unlv.nevada.edu}
\author{F. Javier Aoiz} \affiliation{Departamento de Qu\'{\i}mica F\'{\i}sica. Universidad Complutense.
Madrid 28040, Spain}\email{aoiz@quim.ucm.es}
\date{\today}

\begin{abstract}
%Abstract limit 600 characters. Letter 3750 words.

Cold collisions of light molecules are often dominated by a single partial wave
resonance. For the rotational quenching of HD($v=1,j=2$) by collisions with ground state
para-H$_2$, the process  is  dominated by a single $L=2$ partial wave resonance centered
around 0.1\,K.  Here, we show that this resonance can be switched on or off simply by
appropriate alignment of the HD rotational angular momentum relative to the initial
velocity vector, thereby enabling complete control of the collision outcome.

\end{abstract}

\maketitle

%\section{\label{sec:level1}First-level heading}
% sections are not used for PRL papers

%\section{Introduction}

At cold ($<$ 1~K) and ultracold ($<$ 1~$\mu$K) temperatures molecules can be
prepared in precisely defined quantum states and interrogated with
unprecedented precision. Recent developments in molecule cooling and trapping
technologies~\cite{ wynar.freeland.ea:molecules,regal.ticknor.ea:creation,
sawyer.lev.ea:magnetoelectrostatic,shuman.barry.ea:laser, hummon.yeo.ea:2d,
akerman.karpov.ea:trapping,
anderegg.augenbraun.ea:radio,truppe.williams.ea:molecules} as well as merged or
co-expanding beam
techniques~\cite{henson.gersten.ea:observation,jankunas.bertsche.ea:dynamics,
klein.shagam.ea:directly,perreault.mukherjee.ea:supersonic,suitsJPCL17,naulin2014}
have made it increasingly possible to study molecular systems at these low
temperatures. Such systems have even been used in the frontiers of particle
physics,~\cite{demille.doyle.ea:probing} for example  in the search for the
electric dipole moment of the electron.~\cite{
hudson.kara.ea:improved,baron.campbell.ea:order,
cairncross.gresh.ea:precision} Cold and ultracold molecules therefore offer an
ideal platform on which to precisely study fundamental aspects of molecular
dynamics~\cite{ bell.softley:ultracold,carr.demille.ea:cold,
balakrishnan:perspective,bohn.rey.ea:cold} such as the role of quantum
statistics,~\cite{ospelkaus.ni.ea:quantum-state} threshold law,s~\cite{
balakrishnan.dalgarno:chemistry} and geometric-phase
effects.~\cite{kendrick.hazra.ea:geometric*1}

One of the most fundamental questions in molecular dynamics is the dependence of a
collision outcome on the relative orientation and/or alignment of the reactants -- the
stereodynamics of a collision process.~\cite{ bernstein.herschbach.ea:dynamical,
levine.bernstein:molecular,orr-ewing.zare:orientation, orr-ewing:dynamical,
aoiz.brouard.ea:new,Sharples2018,heidnc2019,LiuNC2012,LiuJCP2014} At cold and ultracold temperatures, where
collisions proceed through just one or a few partial waves, their  stereodynamics  can be
studied at their most fundamental level -- the single quantum state level. In a recent
series of papers Perreault~\etal{} have examined the role that the initial alignment of
HD plays in cold collisions with H$_2$ and D$_2$~\cite{
perreault.mukherjee.ea:quantum,perreault.mukherjee.ea:cold}. Control over rotational
quenching rates was demonstrated, and subsequent theoretical studies revealed that for
certain states the scattering dynamics of cold HD+n-\HH{} collisions is determined by a
single ($L=2$) partial-wave shape resonance at around
1~K~\cite{croft.balakrishnan.ea:unraveling,croft.balakrishnan:controlling}.

While the stereodynamics of atom-diatom collisions has been explored in previous
theoretical
studies~\cite{miranda.clary:quantum,kandel.alexander.ea:cl,jambrina.aldegunde.ea:effects,
aldegunde.javier-aoiz.ea:quantum, jambrina.menendez.ea:how,BrouardJCPexp}, collisions
between oriented and/or aligned molecules in cold conditions remain largely unexplored.
In this work we apply the theoretical methods required to describe the stereodynamics of
inelastic molecule-molecule collisions, specifically, to one of the prototypical systems
recently studied by Perreault~\etal{} -- rotational quenching of HD in cold collisions
with \HH{}. In particular, we will demonstrate how the stereodynamics of cold
molecule-molecule collisions can be determined by a single partial wave shape resonance
and how it can be used  to obtain exquisite control of the collision outcome.

Quantum Mechanical (QM) inelastic scattering calculations were carried out using the
time-independent coupled-channel formalism within the total angular momentum (TAM)
representation of Arthurs and Dalgarno~\cite{ arthurs.dalgarno:theory}, which has
previously been successfully applied to collisions of H$_2$ with
H$_2$~\cite{schaefer.meyer:theoretical,pogrebnya.clary:full-dimensional,quemener.balakrishnan.ea:vibrational}
and HD ~\cite{schaefer:rotational,flower:quantum,balakrishnan.croft.ea:rotational}. The
scattering calculations were performed using a modified version of the TwoBC
code~\cite{krems, quemener.balakrishnan:quantum} on the full-dimensional potential
surface of Hinde~\cite{hinde:six-dimensional}. In the TAM representation the rotational
angular momenta of the dimers, $\bm j_{_{\rm H_2}}$ and $\bm j_{_{\rm HD}}$, are coupled
to form $\bm j_{12} = \bm j_{_{\rm H_2}} + \bm j_{_{\rm HD}}$ which is in turn coupled
with the orbital angular momentum $\bm L$ to form the total angular momentum $\bm J = \bm
L + \bm j_{12}$. Scattering calculations are performed separately for each value of the
total angular momentum $J$ and parity $I=(-1)^{j_{_{\rm H_2}}+j_{_{\rm HD}}+L}$ that
reflects the inversion symmetry of the wavefunction \cite{Alexander77}, yielding the
Scattering matrix, $S^{J}_{\gamma, \gamma'}$, labeled by the asymptotic entrance and exit
channels $\gamma$ and $\gamma'$ respectively. The composite index $\gamma = v_{_{\rm
HD}}\, j_{_{\rm HD}}\, v_{_{\rm H_2}}\, j_{_{\rm H_2}}\, L\, j_{12}$  denotes the
vibrational and rotational quantum numbers of each dimer ($v$ and $j$ respectively), the
orbital angular momentum ($L$), and the sum of the angular momenta of the dimers
($j_{12}$). The integral cross section for state-to-state rovibrationally inelastic
collisions is then given in terms of the S-matrix by
\begin{eqnarray} \nonumber
  \sigma_{\alpha \to \alpha'} &=&  \frac{\pi}{(2j_{_{\rm H_2}}+1)(2j_{_{\rm HD}}+1)k_\alpha^2}
\\ && \sum_{\gamma, \gamma'}(2J+1) |\delta_{\gamma, \gamma'} - S^{J}_{\gamma, \gamma'}|^{2},
\label{eqn:ics}
\end{eqnarray}
where $\alpha$ is the combined molecular state, $\alpha=v_{_{\rm H_2}}\, j_{_{\rm H_2}}\,
v_{_{\rm HD}}\, j_{_{\rm HD}}$, $k^2=2 \mu E_{\rm col}/\hbar^2$ is the square of the wave
vector, $E_{\rm col}$ the collision energy, and $\mu$ the reduced mass. From the elements
of the S-matrix, the scattering amplitudes, $f_{\alpha' \Omega',\alpha \Omega}$ where
$\Omega$ ($\Omega'$) are the helicities, the projection of $j$ ($j'$) into the approach
(recoil) direction,  were determined using the procedure described in
Ref~\citenum{croft.balakrishnan.ea:unraveling}.

Inelastic collisions of HD($v=1, j=2$) with H$_2$($v=0, j=0$) at low collision energies
are dominated by $\Delta j=-1$ and $-2$ transitions  in HD leading to HD($v'=1,
j'=1$)+H$_2$ and HD($v'=1, j'=0$)+H$_2$, respectively. Vibrational de-excitation of HD is
energetically allowed, but  the cross section for vibrational relaxation is around 5-6
orders of magnitude smaller at these collision energies. Energetically, two-quanta
rotational excitation of H$_2$ is  not allowed.
\begin{figure}
  \centering
  \includegraphics[width=1.0\linewidth]{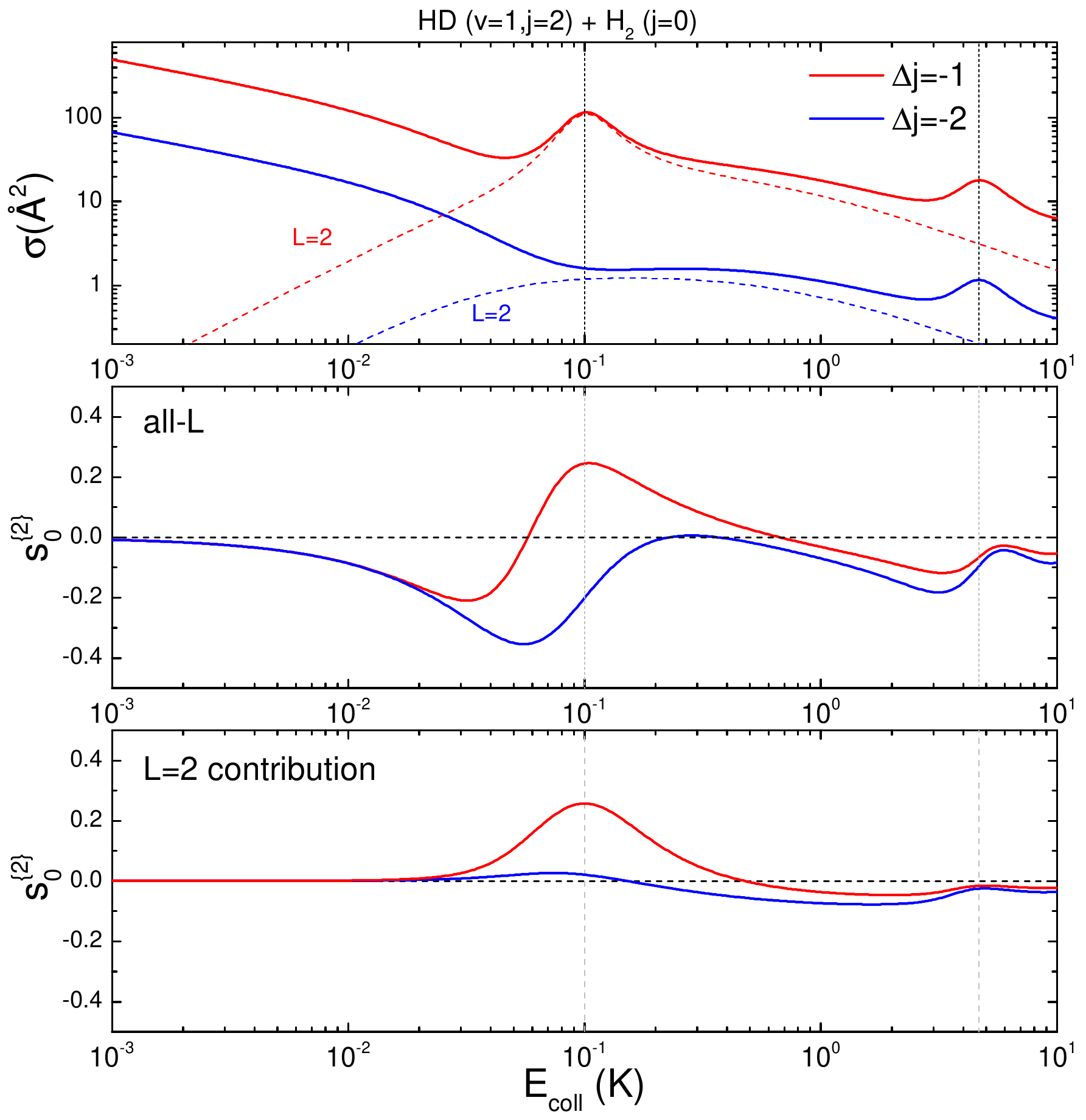}
  \caption{
   Cross sections for the HD($v$=1,$j$=2) + H$_2$($v$=0,$j$=0) inelastic
    collisions as a function of the collision energy. Top panel: cross sections for  $\Delta j=-1$
    (solid red line) and $\Delta j=-2$ (solid blue line).
     The contributions of the $l$=2 partial wave to the cross sections are shown as
    dashed lines. Middle and bottom panels: Energy dependence of the  $s^{\{2\}}_0$
    integral alignment moment for $\Delta j=-1$ (red) and $-2$ (blue).
    In the middle panel the overall results are shown, while the contribution
    for $L=2$ is displayed in the bottom panel.
 }
  \label{Fig1}
\end{figure}

The energy dependence of the rotational quenching cross sections is shown in the top
panel of Fig.~\ref{Fig1}. It is seen that at the lowest energies considered, the cross
section for $\Delta j=-1$ is about a factor of seven larger than for $\Delta j=-2$. Both
show the onset of the Wigner threshold regime below about 0.01~K ($\propto E_{\rm
coll}^{-1/2}$ for pure $s$--wave collisions). The most salient feature for $\Delta j=-1$
is the presence of a sharp resonance at 0.1~K,  where the cross section increases by
almost a factor of four. This is an $L=2$ shape resonance (\Fref{Fig1} shows the
contribution from $L=2$ separately), that is caused by a single S-matrix element
corresponding to $L=2$ and $J=3$ in the TAM representation. As a consequence of this, the
resonance has a defined parity, in this case the block that does not include $\Omega$=0,
which as we will show later has important consequences for the collision mechanism.  This
particular resonance is not observed for $\Delta j=-2$, even though most of the
scattering also comes from $L=2$. Such resonances are ubiquitous features of inelastic
and reactive collisions, especially in the cold regime. Here we show how they can be used
to reveal the collision mechanism and, perhaps more importantly, {\em control the
collision outcome}.

The concept of a collision mechanism can be at times somewhat vague, relying on
qualitative rather than on quantitative results, which can lead to misinterpretations. To
avoid any ambiguities we use the three-vector correlation $\bm k-\bm j_{_{\rm HD}}-\bm
k'$ (where $\bm k$ and $\bm k'$ are the initial and final relative wave vectors) which is
especially well-suited to  characterizing collision mechanisms within a  purely
quantum-mechanical framework. More explicitly  we use the set of reactant polarization
parameters, $s^{\{k\}}_q$, of rank $k$ and component $q=-k \ldots k$, which define the
vector correlation~\cite{AMHKSA:JPCA05}. For the present purposes, the most relevant of
these parameters is $s^{\{2\}}_0$, the first alignment moment of $\bm j$ about the
incoming relative velocity. Its value specifies the direction of the rotational angular
momentum relative to the initial approach. Negative values of $s^{\{2\}}_0$ indicate a
preference for head-on collisions (rotational angular momentum $\bm j_{_{\rm HD}}$
perpendicular to $\bm k$), whereas positive values indicate a preference for side-on
collisions ($\bm j_{_{\rm HD}}$ mostly parallel to $\bm k$). The polarization parameters
are calculated from the integration of the polarization-dependent differential cross
sections, $S^{\{k\}}_q(\theta)$, over the scattering angle, $\theta$. For the $\bm k-\bm
j_{_{\rm HD}}-\bm k'$ correlation the $S^{\{k\}}_q(\theta)$ can be determined from
$f_{\alpha' \Omega',\alpha \Omega}$:\cite{aldegunde.miranda.ea:how}
\begin{eqnarray} \nonumber
S^{(k)}_q(\theta) &=& \frac{1}{2j_{_{\rm HD}}+1} \sum_{\Omega_1 \Omega_2} \sum_{\Omega'}
f_{\alpha' \Omega',\alpha \Omega_1} \left[ f_{\alpha' \Omega',\alpha \Omega_2} \right]^* \\
 && \times \langle j_{_{\rm HD}} \Omega_1, k q | j_{_{\rm HD}} \Omega_2 \rangle
\end{eqnarray}
where $\langle ..,..|..\rangle$ denotes the Clebsch-Gordan coefficient. In the particular
case of $q$=0, the $s^{\{k\}}_q$ can also be obtained directly from the S-matrix elements
\cite{Aldegundejpcl2012}.

The middle panel of~\Fref{Fig1} shows the polarization moment $s^{\{2\}}_0$ as a function
of the collision energy for both the $\Delta j=-1$ and $-2$ transitions. At the lowest
energies,  the moment goes to zero, as required for ultracold
collisions~\cite{AAMSA:JCP06}. With increasing collision energy, s$^{\{2\}}_0$ takes
negative values for both transitions, showing a preference for head-on encounters.
However, at the proximity of the resonance, $s^{\{2\}}_0$ exhibits markedly different
behavior for the two transitions. It turns positive for $\Delta j=-1$ peaking at the
energy of the resonance (denoted with a vertical dashed line), while for $\Delta
j=-2$ it remains negative. This shows that the resonance for $\Delta j=-1$ is associated
with a specific mechanism that is not shared by the $\Delta j=-2$ transition. At energies
above the resonance, s$^{\{2\}}_0$ again shows the same trend for both transitions, with
a small change around 4.75\,K caused by a second resonance (present in both $\Delta j$
transitions) that does not change the mechanism significantly.

To unambiguously analyze the effect of the resonance, the $L=2$ contribution to
$s^{\{2\}}_0$ is shown in the bottom panel of~\Fref{Fig1}. It is calculated by including
only the $L=2$ elements of the scattering matrix (without considering their coherence
with other $L$ values). Regardless of $\Delta j$, the $L=2$ contribution to $s^{\{2\}}_0$
goes to zero at ultracold energies, as does $\sigma_{L=2}$ (see top panel). Moreover, up
to 0.5~K, including the resonance, the sign of the $L=2$ contribution  to $s^{\{2\}}_0$
is positive (favoring side-on collisions) while it is negative for higher collision
energies. Although at the resonance the sign of the $L$=2 contribution to  $s^{\{2\}}_0$
is positive for both $\Delta j$=-1 and -2, its magnitude is much larger for the former.
Since, $L=2$ collisions dominate around 0.1~K for both $\Delta j$ (see top panel of
Fig.~\ref{Fig1}), these results indicate that the overall change of $s^{\{2\}}_0$, and
hence of the collision mechanism, is caused by the resonance and not due to a larger
contribution of $L=2$.
\begin{figure}
  \centering
  \includegraphics[width=1.0\linewidth]{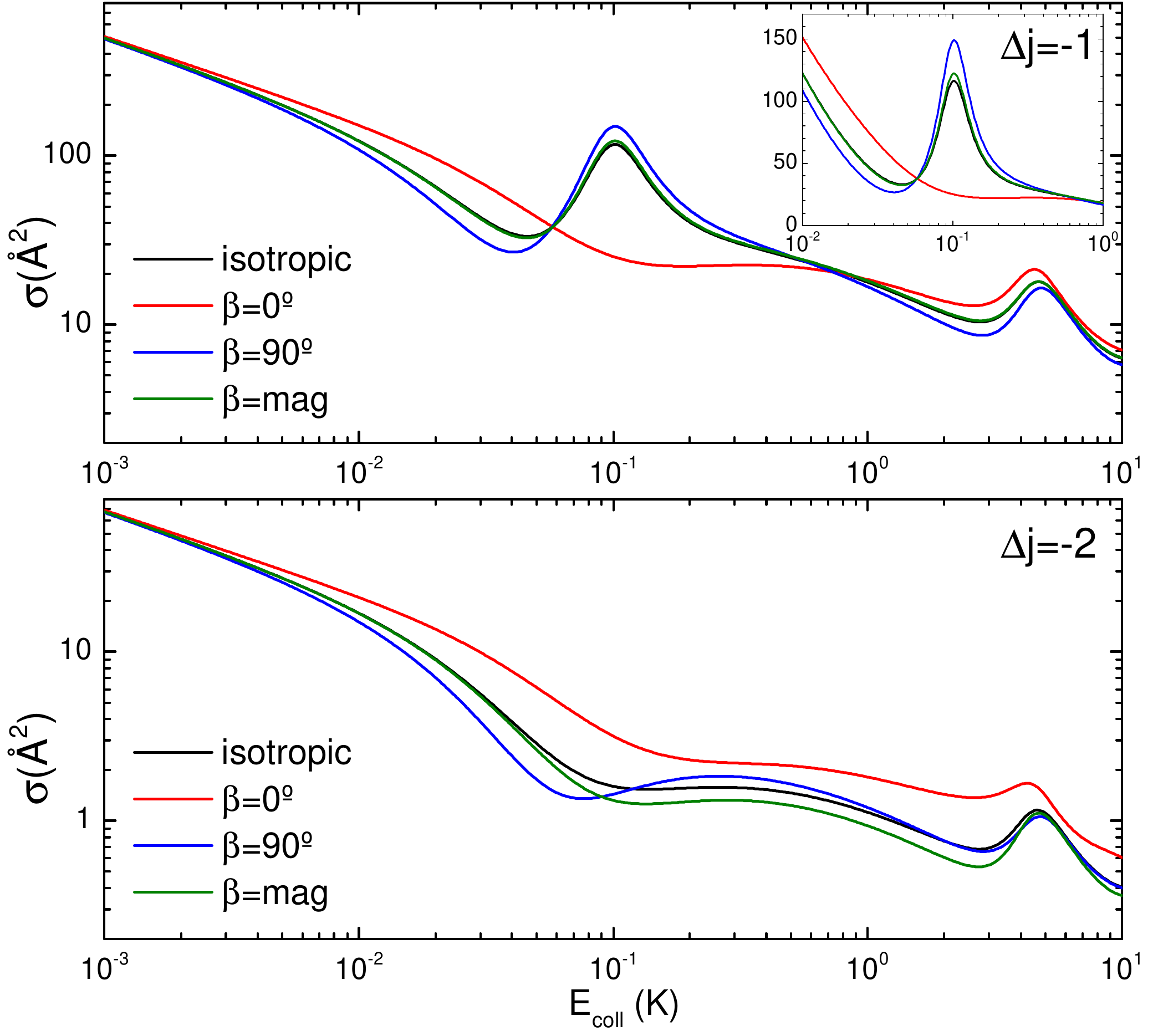}
  \caption{Top panel, cross sections as a function of the collision energy
  for $\Delta j=-1$ for different preparations of the HD
    internuclear axis, $\beta=0^{\circ}$ (red line), $\beta=90^{\circ}$
    (blue line), and the magic angle (olive line). The isotropic preparation
    (in absence of external alignment) is shown in black.
    The inset shows a zoom of the resonance region in a linear ordinate-axis scale.
    Bottom panel: Same as top panel but for $\Delta j=-2$.}\label{Fig2}
\end{figure}

The distinct mechanism for the resonance suggests that it might be possible to suppress
its effect by  appropriate state-preparation of the HD rotational angular
momentum~\cite{perreault.mukherjee.ea:quantum, perreault.mukherjee.ea:cold}. The cross
sections for different extrinsic preparations can be computed following the procedure
described in Ref.~\citenum{aldegunde.miranda.ea:how}. If HD is prepared in a directed
state,  $m$=0, where $m$ is the magnetic quantum number, it leads to the alignment of the
internuclear axis along the quantization axis (in the case of
refs.~\citenum{perreault.mukherjee.ea:quantum,perreault.mukherjee.ea:cold} the
polarization vector of the pump and Stokes lasers). By varying the direction of the
laboratory-fixed axis with regard to the scattering frame it is possible to change the
external preparations generating different relative geometries of the reactants prior the
collision. We will label the different extrinsic preparations using $\beta$ and $\alpha$,
where $\beta$ is the polar angle between the polarization vector and the initial relative
velocity, and $\alpha$ is the azimuthal angle that defines the direction of the
polarization vector with respect to the $\bm{k-k'}$ frame. Accordingly, $\beta=0^{\circ}$
implies head-on collisions while $\beta=90^{\circ}$ involves side-on collisions. The
equations that relate the observed differential cross section (DCS) for a given
preparation ($d \sigma_{\alpha}^{\beta} / d \omega$) is given
by:\cite{aldegunde.miranda.ea:how}
\begin{equation}\label{dcsalphabeta}
\frac{d \sigma_{\alpha}^{\beta}}{d \omega}  = \sum_{k=0}^{2j_{_{\rm HD}}} \sum_{q=-k}^{k} (2 k + 1 )
\left[S^{(k)}_q(\theta)\right]^* A^{(k)}_0  C_{kq}(\beta,\alpha)
\end{equation}
where $C_{kq}(\beta,\alpha)$ are the modified spherical harmonics and extrinsic moments
$A^{(k)}_q$ that defined the preparation in the laboratory frame are derived in
Ref~\citenum{aldegunde.miranda.ea:how}. The integral cross section can be obtained by
integrating $d \sigma_{\alpha}^{\beta} / d \omega$ over the scattering and the azimuthal
angles, hence depending only on $\beta$.

Fig.\ \ref{Fig2} shows the cross sections for different experimentally achievable
extrinsic preparations of the HD rotational angular momentum for an unpolarized H$_2$
molecule. The results for $\Delta j=-2$ are relatively featureless, and are identical to
those shown in Ref.~\citenum{croft.balakrishnan:controlling}. In the Wigner threshold
regime, no control can be attained for the integral cross section~\cite{AAMSA:JCP06}.
With increasing collision energy, however, $\beta=0^{\circ}$ always leads to larger cross
sections (by up to a factor of 2). The effect of $\beta=90^{\circ}$, and $\beta=$~mag
(magic angle) preparations is milder, leading to only small changes in the cross sections
with respect to the unpolarized case.

For $\Delta j=-1$ the situation is similar for energies below the resonance. However, at
the resonance the collision mechanism changes rather abruptly, and the $\beta=0^{\circ}$
preparation, which implies head-on collision, leads to a sudden decrease of the cross
section, by close to a factor of 5, the most extreme effect that could be observed for
any preparation of a sharp $j=2$ state.  Since the $\beta=0^{\circ}$ preparation is the
same as collisions with $\Omega=0$ exclusively, the fact that the S-matrix element that
causes the resonance does not contain $\Omega=0$ leads to the disappearance of the
resonance.

Well above the resonance, at $E_{\rm coll}\ge$0.6~K, the effect somewhat reverts back to
the behavior observed below the resonance, with $\beta=0^{\circ}$ again leading to a
slight increase in the cross section. To sum up, the alignment of $\bm j_{_{\rm HD}}$
perpendicular to $\bm k$ slightly enhance the cross section except at the resonance,
where it brings about the suppression of the resonance as if it were switched off. The
effect of other preparations $\beta=90^{\circ}$ and $\beta=$~mag is relatively minor and,
apparently, does not affect the resonance significantly, as far as the integral
cross section is concerned.
\begin{figure}
  \centering
  \includegraphics[width=1.0\linewidth]{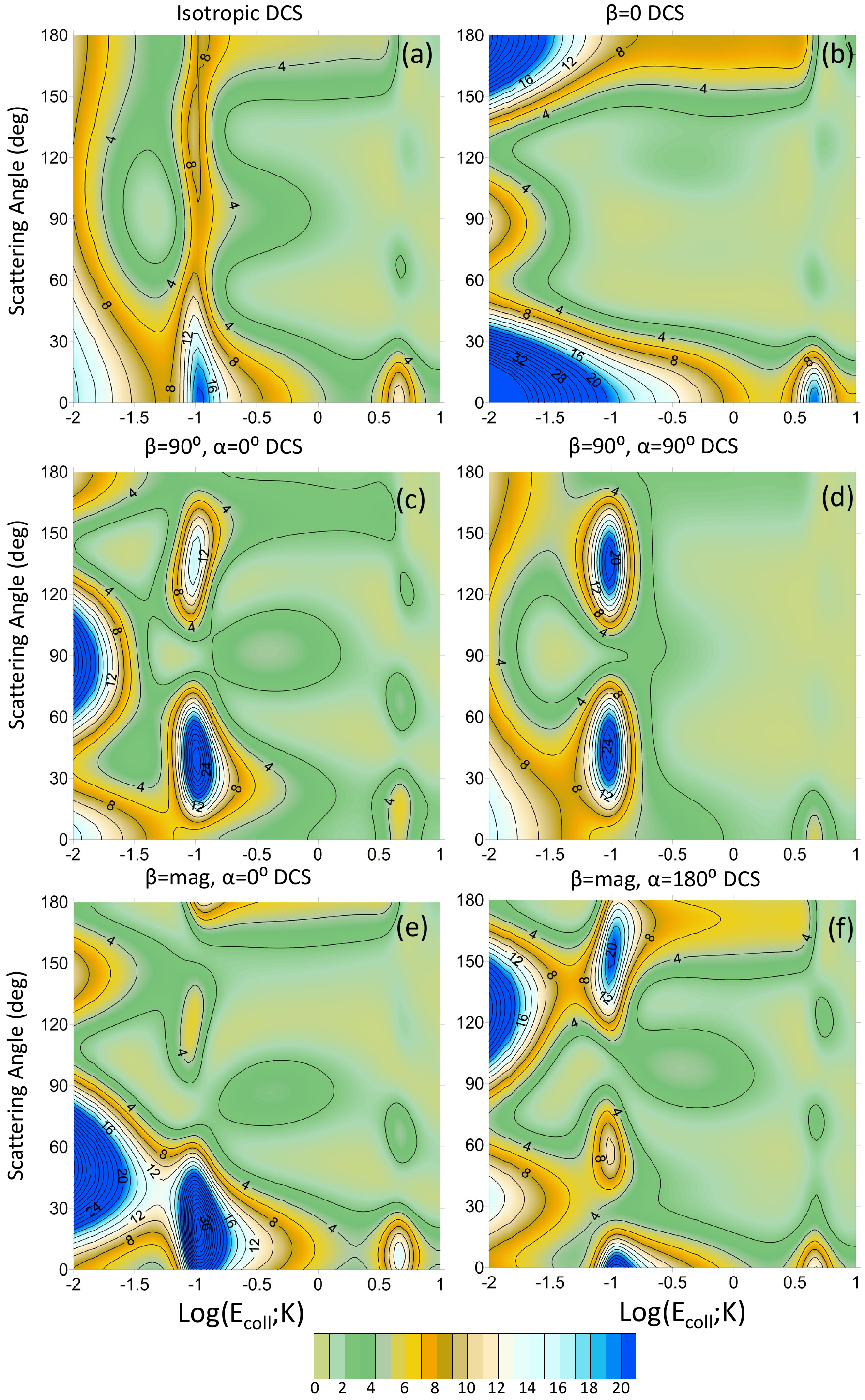}
  \caption{
    Contour plots showing the collision energy dependence of the DCS for the $\Delta j=-1$
    transition with different preparation of the HD rotational angular momentum.
    The angles $\beta$ and $\alpha$ are the polar and azimuthal angles,
    respectively, that define the direction of the light polarization vector with
    respect to the $\bm k$-$\bm k'$ scattering frame (where $\bm k$ defines the
    $z$ axis, and the reference plane contains $\bm k$ and $\bm k'$, where $\bm k'$ is the recoil
    velocity). While the integral cross sections at a given energy depends only on $\beta$,
    the DCS depends on both angles. The effect of the resonance is prominent for
    all preparations except for $\beta$=0, for which the resonance disappears.
    }\label{Fig3}
\end{figure}

Up to this point, we have shown that at the resonance there is a change in the collision
mechanism, which can be used to control the cross section by changing the
preparation of the HD rotational angular momentum. It has been demonstrated recently by
Perreault~\etal{} that it is possible to determine the DCS for different reagent
preparations~\cite{ perreault.mukherjee.ea:quantum,perreault.mukherjee.ea:cold}, so we
now shift our attention to investigating how the DCS is affected by state-preparation of
the HD molecule. Fig.~\ref{Fig3} shows the DCS as a function of the scattering angle and
collision energy for $\Delta j=-1$. The isotropic DCS (with unpolarized collision
partners) is shown in panel (a), which features a slight preference for forward
scattering. In particular, the resonance appears as a sharp ``ridge'' with a  clear
preference for forward scattering. For $\beta=$0$^{\circ}$, panel (b), the situation is
completely different. First, the resonance completely vanishes, and at 0.1~K there are no
marked changes or discontinuities in the energy dependence of the DCS. In addition, the
shape of the DCS displays prominent forward and backward peaks irrespective of the
collision energy. At low collision energies there is a third peak in the DCS that only
survives for energies below 0.03~K. There is also a resonance at higher collision
energies, around $E_{\rm coll} \sim$ 5~K, which unlike the 0.1~K resonance is slightly
enhanced by this external preparation.

While the $\beta=90^{\circ}$ and $\beta=$~mag preparations have a minor effect on the
integral cross section, the polarization of $\bm{j}_{\rm HD}$ has  a dramatic effect on
the shape of the DCS. Fig.~\ref{Fig3} (panels (c)--(f)) shows the effect of
$\beta=90^{\circ}$ and $\beta=$~mag and $\alpha=0^{\circ}, 180^{\circ}$ preparations on
the DCS. The shape and magnitude of the DCS for all these cases differ from each other
and from the isotropic case. Moreover, all of them show distinct features at the
resonance. For $\beta=90^{\circ}$, $\alpha=0,90^{\circ}$, the DCS at the resonance has
two prominent peaks at around $30^{\circ}$ and $150^{\circ}$, while for $\beta=$~mag and
$\alpha=0^{\circ}$ there is a strong enhancement of forward scattering at the resonance.
While for all non-zero $\beta$ values the resonance at 0.1~K is present, its angular
distribution is strongly sensitive to $\beta$, showing that the resonance can be used to
control not just the magnitude of the cross section, but also the scattering direction.
\cite{heidnc2019}  This provides a powerful tool to elucidate stereodynamics of
resonance-mediated collisions and fine-tune calculated interaction potentials against
controlled experiments.

\begin{figure}
\centering
\includegraphics[width=1.0\linewidth]{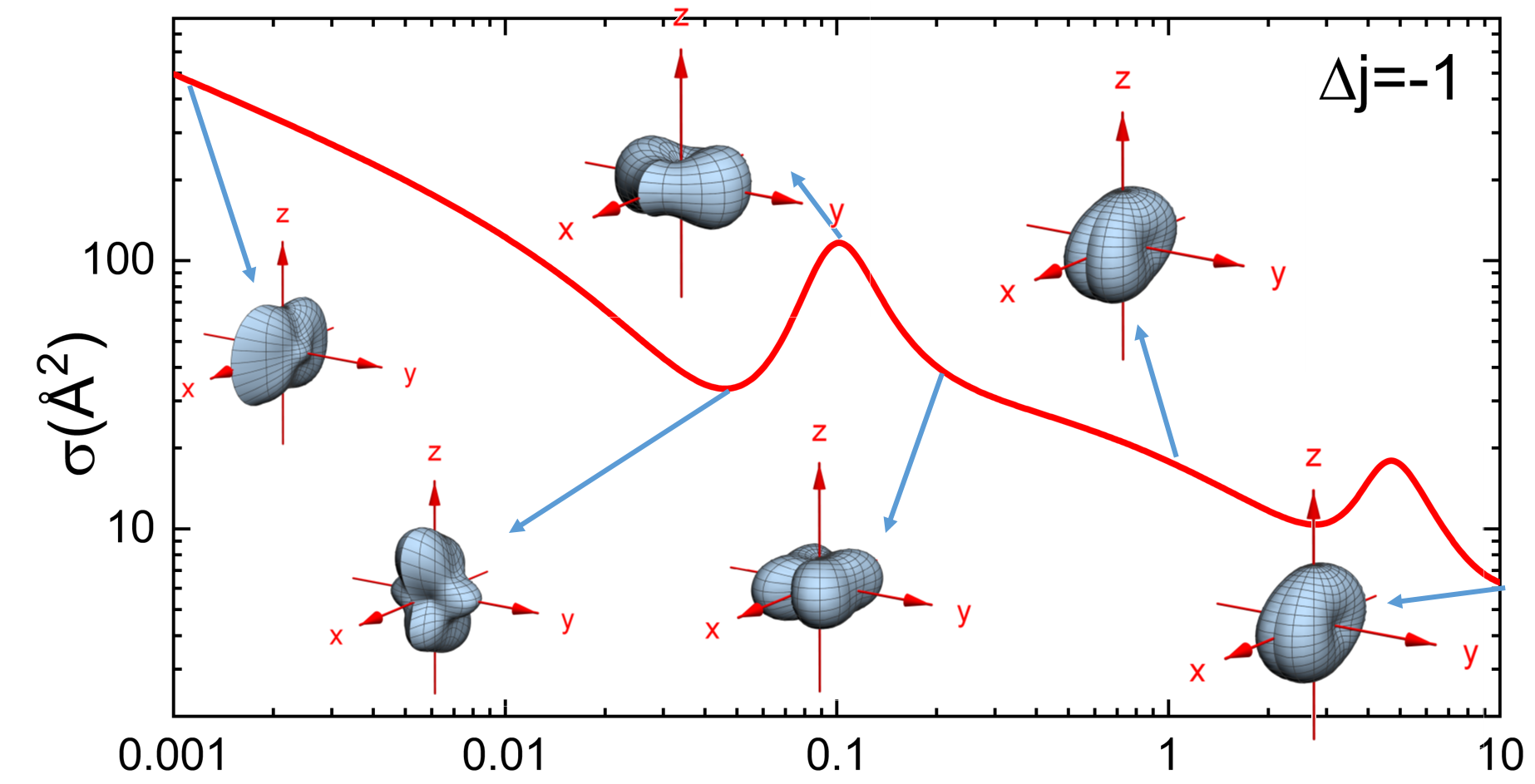}
\includegraphics[width=1.0\linewidth]{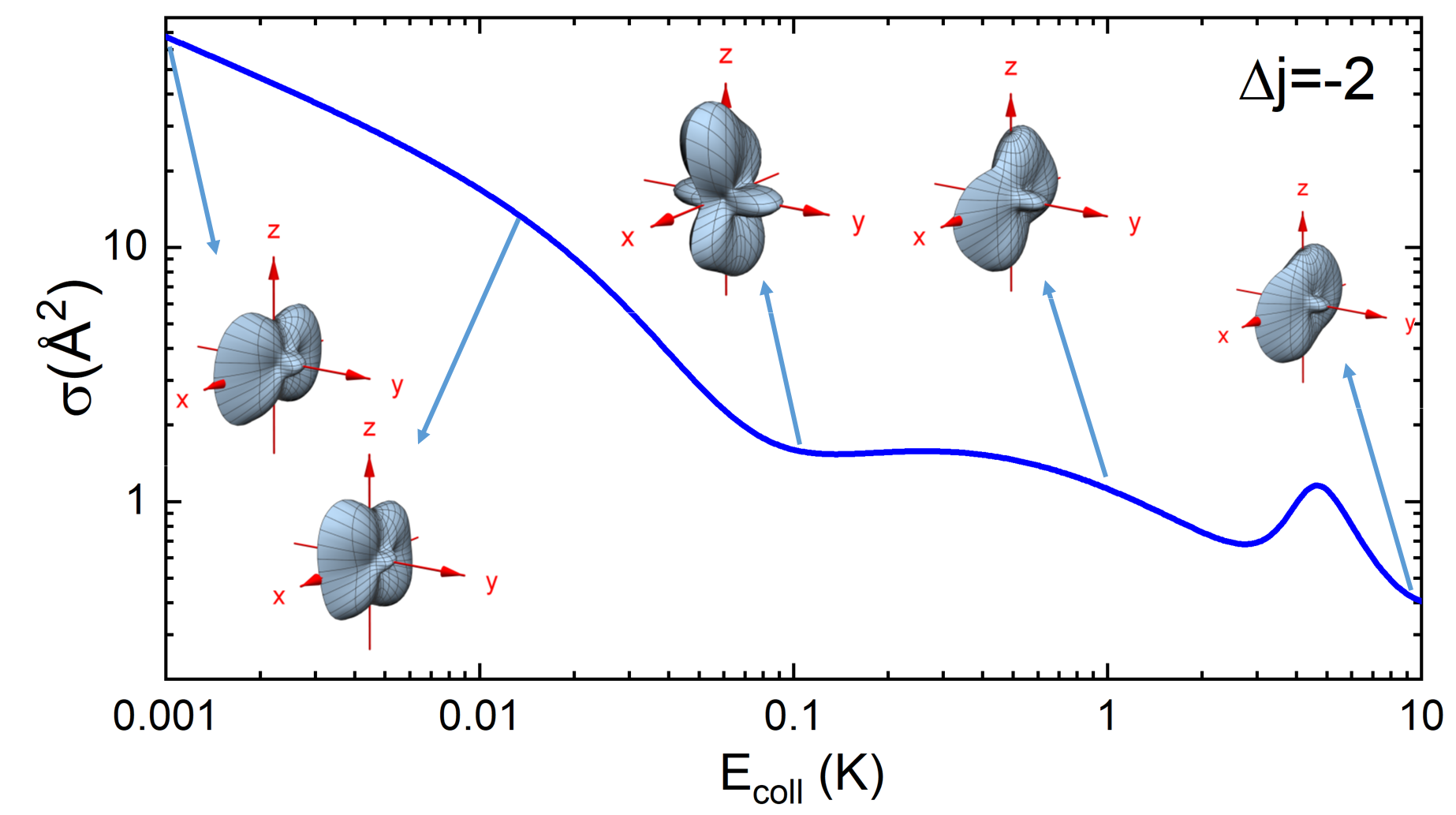}
\caption{
  Cross sections as a function of the collision energy for $\Delta j=-1$ (top panel) and $-2$ (bottom
  panel) along with the internuclear axis stereodynamical portraits (see text for
  further details) that reveal the change of mechanism at the resonance. The
  reference frame for the stereodynamical portraits is defined by the reactants
  approach ($\bm{k}$) and the products recoil ($\bm{k'}$ ) directions.  The $z$ axis is
  parallel to $\bm{k}$, the $x$-$z$ plane is the scattering plane, and $y$ axis
  is parallel to  $\bm{k}$ $\times$ $\bm{k'}$.
  }\label{Fig4}
\end{figure}

To gain further insight into the reaction mechanism we analyze the remaining polarization
parameters besides s$^{\{2\}}_0$. For initial $j=2$ there are 12 independent parameters
of which 8 contribute to the alignment of the internuclear axis distribution. To aid the
interpretation of these parameters we use ``stereodynamical
portraits''~\cite{miranda.aoiz:interpretation,miranda.aoiz:interpretation2}, which show
the spatial distribution of the angular momentum on the internuclear axis for a given
polarization of the rotational angular momentum, with the $z$ axis parallel to $\bm{k}$.
Top panel of figure~\ref{Fig4} presents the stereodynamical portraits associated with the
internuclear axis of HD for $\Delta j=-1$. At the lowest energies considered here, the HD
internuclear axis is contained in the scattering plane, although it does not show a
significant preference towards head-on or side-on encounters. Just below the resonance,
however, it starts to show a strong preference towards head-on collisions (typically
associated with small impact parameters). A sudden change of the mechanism occurs at the
resonance, with a clear preference for side-on encounters (internuclear axis
perpendicular to $z$). Just above the resonance the internuclear axis remains
perpendicular to the approach direction, but preferentially contained in the $xy$
plane. With increasing collision energy, the internuclear axis is no longer aligned
along or perpendicular to $z$. For $\Delta j =-2$, the stereodynamical portraits are
similar except at the resonance found for $\Delta j=-1$ where, in contrast, the head-on
encounters are preferred, evincing the dramatic change of the reaction mechanism caused
by the resonance.

Altogether, these results demonstrate that, in the cold-energy regime, inelastic
collisions between HD($v$=1,$j$=2) and p-H$_2$ are controlled by a resonance at 0.1~K
that causes profound changes to the reaction mechanism that favors side-on collisions,
typically associated with large impact parameters, over head-on collisions that would
have been preferred if the resonance were absent. This sudden change in mechanism permits
exquisite control of the collision outcome by using different preparations of the HD
internuclear axis, and makes it possible to switch-off the resonance altogether. The
effect of the initial HD alignment becomes most evident in the DCS, which changes
dramatically for the alternative preparations investigated. Energy resolved measurements
of angular distribution of state-prepared HD in collisions with H$_2(v=0, j=0)$ would be
desirable to validate these predictions.

P.G.J. acknowledges funding by the Fundaci\'on Salamanca city of culture and knowledge
(programme for attracting scientific talent to Salamanca).  J.F.E.C. gratefully
acknowledges support from the Dodd-Walls Centre for Photonic and Quantum Technologies. H.
G. acknowledges US Department of energy (DE-SC0015997).  M.B. thanks support of the UK
EPSRC (to M.B. via Programme Grant EP/L005913/1). N.B. acknowledges support from the US
National Science Foundation, Grant No. PHY-1806334. P.G.J. and F.J.A. acknowledges
funding from the Spanish Ministry of Science and Innovation (Grants No. CTQ2015-65033-P
and PGC2018-09644-B-100).

\bibliographystyle{apsrev4-1}

\bibliography{sterodynamics,h4}

\end{document}